\documentstyle[12pt]{article}

\begin{document}

\author{C. S. Unnikrishnan \\
{\it Gravitation Group, Tata Institute of Fundamental Research, }\\
{\it Homi Bhabha Road, Mumbai - 400 005, INDIA},\\
and\\
{\it NAPP Group, Indian Institute of Astrophysics,}\\
{\it Koramangala, Bangalore - 560 034, INDIA}}
\title{There is no spooky action-at-a-distance in quantum correlations: Resolution
of the Einstein-Podolsky-Rosen nonlocality puzzle}
\maketitle

\begin{abstract}
The long-standing puzzle of the nonlocal Einstein-Podolsky-Rosen
correlations is resolved. The correct quantum mechanical correlations arise
for the case of entangled particles when strict locality is assumed for the
probability amplitudes instead of locality for probabilities. Locality of
amplitudes implies that measurement on one particle does not collapse the
companion particle to a definite state.
\end{abstract}

Sixty five years ago, the most significant paper questioning a fundamental
aspect of quantum phenomena was written by Einstein, Podolsky and Rosen
(EPR) \cite{epr}. They addressed the question whether the wave-function
represented a complete description of reality in quantum mechanics, and
argued that it didn't. Bohr's reply to this paper \cite{bohr} was not
sufficient to resolve the fundamental issues raised by EPR. Decades later,
Bohm rephrased the EPR problem \cite{bohm} in terms of particles correlated
in their spin and this helped enormously in analyzing the problem with
clarity. John Bell analyzed the EPR problem in the early sixties and
established the Bell's inequalities obeyed by any local hidden variable
theory for the correlations of entangled particles \cite{bell}. Quantum
mechanical correlations calculated using the entangled wave-function and
spin operators violate these inequalities. Experiments, the first of which
was by Freedman and Clauser \cite{clauser} and the most remarkable by A.
Aspect and collaborators \cite{aspect}, have established beyond doubt that
there cannot be a viable local realistic hidden variable description of
quantum mechanics \cite{debate}. Further, these results also have been
interpreted as evidence for nonlocal influences in quantum measurements
involving entangled particles. Since no instruction set carried by the
particles from their source of origin (possibly with the addition of several
local hidden variables) can manage to create the correct correlations
observed in experiments, the only way out seems to be that measurement of an
observable on one of the particles in an entangled pair seems to convey the
result of this measurement instantaneously to the other particle resulting
in the correct behaviour of the other particle during a measurement on the
second particle. Of course, the no signalling theorems in this context
prohibit any faster than light signalling using this feature. Nevertheless,
we seem to be stuck with the puzzling nonlocality which is probably the
deepest mystery in the behaviour of entangled systems. In the quantum
mechanical terminology, the measurement of an observable on one of the
particles collapses the entire wave-function instantaneously and nonlocally
and the second particle acquires a definite value for the same observable,
consistent with the correlation determined by the relevant conservation law.

Apart from the disturbing aspect of accepting the concept of nonlocality
without being able to understand its nature, there is serious conflict with
the spirit of relativity. As soon as we bring in the concept of one
measurement being influenced nonlocally by the other, the notion of
simultaneity becomes important since both measurements can be labelled by
local times. So, if one measurement precede the other in one frame, one can
always find a moving frame in which the converse it true, the second
measurement preceding the first \cite{penrose}.

In this paper we discuss the resolution of the quantum nonlocality puzzle.
The crucial new idea is to assume locality at the level of probability
amplitudes instead of at the level of probabilities. For quantum systems
which show wave-like behaviour represented by complex numbers, this seems to
be the physically correct assumption to make. The quantum correlation is
encoded in the difference of an internal variable for the problem.

Consider the breaking up of a correlated state as in the standard Bohm
version of the EPR problem \cite{bohm}. The two particle go off in opposite
directions and are in space-like regions. Two observers make measurements on
these particles individually at space like separated regions with time
stamps such that these results can be correlated later through a classical
channel. We assume that strict locality is valid at the level of probability
amplitudes. A measurement changes probability amplitudes only locally. {\em %
Measurements performed in one region do not change the magnitude or phase of
the complex amplitude for the companion particle in a space-like separated
region. }

We assign local rules (probability amplitudes) for the outcome of a
particular measurement on each of the two particles. We also assume the
existence of an internal variable for each of these two particles. The
correlation at source is encoded in the relative value, or the difference,
of this internal variable for the two particles. For simplicity let us call
this internal variable a ``phase'', $\phi $. Note that it is not a dynamical
phase evolving as the particle propagates. It is an internal variable whose
difference (possibly zero) remains constant for the particles of the
correlated pair. The value of $\phi $ can vary from particle to particle,
but the relative phase between the two particles in all correlated pairs is
constant. Consider $\phi $ as a reference for the particles to determine the
angle of a polarizer or analyzer encountered on their way, {\em locally} (we
use the terms polarizer and analyzer in a generic way. They could be
Stern-Gerlach like analyzers for spin $1/2$ particles). The first particle
encounters analyzer \#1 kept at an angle $\theta _{1}$ with respect to some
global direction. We denote this angle of the analyzer with reference to $%
\phi $ as $\theta .$ Similarly, the second particle which has the internal
phase angle $\phi +\phi _{o},$ where $\phi _{o}$ is a constant, encounters
the second analyzer oriented at angle $\theta _{2}$ at another space-like
separated point. Let the orientation of this analyzer with respect to the
internal phase angle of the second particle is $\theta ^{\prime }.$ We have $%
\theta -\theta ^{\prime }=\theta _{1}-\theta _{2}+\phi _{o}.$ (The constant $%
\phi _{o}$ characterizes the correlation.)

An experiment in which each particle is analyzed by orienting the analyzers
at various angles $\theta _{1}$ and $\theta _{2}$ is considered next. At
each location the result is two-valued denoted by ($+1$) for transmission
and ($-1$) for absorption of each particle, for any angle of orientation.
The classical correlation function $P({\bf a},{\bf b})=\frac{1}{N}\sum
(A_{i}B_{i})$ satisfies $-1\leq P({\bf a},{\bf b})\leq 1.$ Here $({\bf a},%
{\bf b})$ denotes the two directions along which the analyzers are oriented
and $A_{i}$ and $B_{i}$ are the two valued results. The Bell correlation $P(%
{\bf a},{\bf b})$ denotes the average of the quantity ({\em number of} {\em %
detections in coincidence }$-${\em \ number of detections in anticoincidence}%
), where `coincidence' denotes both particles showing same value for the
measurement and `anticoincidence' denotes those with opposite values. Our
aim is to calculate the Bell correlation from our formalism employing local
amplitudes and compare it with the quantum mechanical prediction obtained
from the nonlocal entangled wave function and spin operators.

We specify the {\em local rule} {\em for transmission as a complex number}, 
{\em whose square gives the probability of transmission}. The complex
amplitude associated with particle \#1 is $C_{1}=\frac{1}{\sqrt{2}}\exp
(i\theta s)$ for measurements at analyzer \#1, and for particle \#2 is $%
C_{2}=\frac{1}{\sqrt{2}}\exp (i\theta ^{\prime }s)$ at analyzer \#2. (For
the maximally entangled particles, the amplitude for the alternate outcome
at the analyzer differs only by a phase). In these expressions, the quantity 
$s$ is the spin (in units of $\hbar $) of the particle, $1$ for photons and $%
\frac{1}{2}$ for spin-$\frac{1}{2}$ particles. The locality assumption is
strictly enforced since the two complex functions depend only on local
variables and on an internal variable determined at source and then
individually carried by the particles without any subsequent communication
of any sort. The probabilities for the outcomes of measurements at each end
are now correctly reproduced, for any angle of orientation. These
probabilities are $Re(C_{1}C_{1}^{*})=Re(C_{2}C_{2}^{*})=\frac{1}{2}.$ The
correlation function for amplitudes is of the form $Re(C_{1}C_{2}^{*}).$ The
correlation amplitude for an outcome of either $(++)$ or $(--)$ of two
maximally entangled particles is 
\begin{eqnarray}
U(\theta ,\theta ^{\prime }) &=&Re2(C_{1}C_{2}^{*})=Re\{\exp is(\theta
-\theta ^{\prime })\}  \nonumber \\
&=&\cos \{s(\theta -\theta ^{\prime })\}=\cos \{s(\theta _{1}-\theta
_{2})+s\phi _{o}\}.
\end{eqnarray}

We rewrite this as $U(\theta _{1},\theta _{2},\phi _{o})$ since all
references to the individual values of the hidden variable $\phi $ has
dropped out. The square of $U(\theta _{1},\theta _{2},\phi _{o})$ is the
probability for coincidence detection of the two particles through the
analyzers kept at angles $\theta _{1}$ and $\theta _{2}$. ( A distinction is
made between what the quantum system uses as a rule for transmission, and
what we can measure after the act of transmission. The correlation function
is analogous to the two-point amplitude correlations of two independent
electromagnetic fields).

Next we calculate the Bell correlation function $P({\bf a},{\bf b})$ from
the correlation function $U(\theta _{1},\theta _{2},\phi _{o}).$ Since $%
U^{2}(\theta _{1},\theta _{2},\phi _{o})$ is the probability for a
coincidence detection ($++$ or $--$), the quantity $(1-U^{2}(\theta
_{1},\theta _{2},\phi _{o}))$ is the probability for an anticoincidence
(events of the type $+-$ and $-+$). Since the average of the quantity
(number of coincidences $-$ number of anticoincidences) = 
\begin{equation}
U^{2}(\theta _{1},\theta _{2},\phi _{o})-(1-U^{2}(\theta _{1},\theta
_{2},\phi _{o}))=2U^{2}(\theta _{1},\theta _{2},\phi _{o})-1,
\end{equation}
the correspondence between $P({\bf a},{\bf b})$ and $U(\theta _{1},\theta
_{2},\phi _{o})$ is given by the general expression, 
\begin{equation}
P({\bf a},{\bf b})=2U^{2}(\theta _{1},\theta _{2},\phi _{o})-1
\end{equation}

Let us consider for discussion, the case of a correlated state of photons
breaking up into {\em orthogonal polarization states}. This means that if
one photon is transmitted through an analyzer on one side, the other one
will not transmitted for the same orientation of the analyzer on the other
side. So, perfect anti-correlation is implied for $\theta _{1}-\theta _{2}=0$%
. The Bell correlation calculated from quantum mechanics for this case is
given by $-\cos (2($ $(\theta _{1}-\theta _{2})).$ That is, if the analyzers
are oriented at a relative angle of $\pi /2,$ perfect correlation is
obtained. When the relative angle is $\pi /4,$ the quantum mechanical
correlation defined in the Bell way is zero, since there are as many
coincidences as anticoincidences.

The correlation function we derived give, for the case of the photons
discussed above, 
\begin{equation}
U(\theta _{1},\theta _{2},\phi _{o})=\cos \{(\theta _{1}-\theta _{2})+\phi
_{o}\}
\end{equation}

We set $\phi _{o}=\pi /2$ for denoting the correlation of the two orthogonal
photons at source . Then we get 
\begin{eqnarray}
U(\theta _{1},\theta _{2},\phi _{o}) &=&\cos \{(\theta _{1}-\theta _{2})+\pi
/2\}  \nonumber \\
&=&-\sin (\theta _{1}-\theta _{2})
\end{eqnarray}

The probability for coincidence detection is 
\begin{equation}
U^{2}(\theta _{1},\theta _{2},\phi _{o})=\sin ^{2}(\theta _{1}-\theta _{2})
\end{equation}

Correspondingly, the probability for anticoincidence is $1-\sin ^{2}(\theta
_{1}-\theta _{2}).$

We get for the Bell correlation, 
\begin{equation}
P({\bf a},{\bf b})=2\sin ^{2}(\theta _{1}-\theta _{2})-1=-\cos (2((\theta
_{1}-\theta _{2}))
\end{equation}

This agrees completely with the usual quantum mechanical prediction derived
by applying the relevant spin operators on the correct entangled state of
the two photons.

Another important example is the case of the singlet state breaking up into
two spin $1/2$ particles propagating in opposite directions to spatially
separated regions. We set $\phi _{o}=\pi .$ Then our correlation function is 
\begin{eqnarray}
U(\theta _{1},\theta _{2},\phi _{o}) &=&\cos \{s(\theta _{1}-\theta
_{2})+s\phi _{o}\}  \nonumber \\
&=&\cos \{\frac{1}{2}(\theta _{1}-\theta _{2})+\pi /2\}  \nonumber \\
&=&-\sin \frac{1}{2}(\theta _{1}-\theta _{2})
\end{eqnarray}

The probability for joint detection through two Stern-Gerlach analyzers
oriented at relative angle $\theta _{1}-\theta _{2}$ is 
\begin{equation}
U^{2}(\theta _{1},\theta _{2},\phi _{o})=\sin ^{2}(\frac{1}{2}(\theta
_{1}-\theta _{2}))
\end{equation}

For the case of the two particles of the singlet state, 
\begin{eqnarray}
2U^{2}(\theta _{1},\theta _{2},\phi _{o})-1 &=&2\sin ^{2}(\frac{1}{2}(\theta
_{1}-\theta _{2}))-1  \nonumber \\
&=&-\cos (\theta _{1}-\theta _{2})=-{\bf a}\cdot {\bf b}
\end{eqnarray}

This is again exactly same as the correct Bell correlation $P({\bf a},{\bf b}%
)$ for the quantum mechanical predictions obtained from the singlet
entangled wave-function and the Pauli spin operators. Perfect correlation is
obtained for oppositely oriented analyzers and perfect anticorrelation for
similarly oriented analyzers. When the analyzers are orthogonal, the
correlation is zero.

We have correctly reproduced the quantum mechanical correlation using local
probability amplitudes. Bell's theorem prohibiting local realistic theories
is not violated since we used the concept of locality for probability
amplitudes instead of locality at the level of probabilities. The correct
correlation emerges from combining two local complex functions. Single
events consisting of two independent measurements at the two analyzers obey
the correlation we derived, and the probability for joint detection is given
by the square of the correlation function. {\em In particular if the two
analyzers happen to be in the same orientation, perfect correlation is
reproduced every time within the strict locality assumption}. It is
important to note that we have not used any information on the internal
variable $\phi $ even in terms of distributions. It may be considered as a
hidden variable appearing in the measurement prescriptions only through a
complex number and has the nature of the origin of a non-dynamical phase
associated with the quantum system. In fact, such a variable is not an
external input additional to what is already available in the quantum
mechanical description, since the zero of the phase of a wave-function is
unobservable.

All probabilities are guaranteed to be positive definite in our formalism
since the correlation function is real. The nonlocality puzzle in the EPR
correlations is resolved. Strict locality including Einstein locality is
valid. An answer to the EPR query regarding the completeness of quantum
description is found. {\em It seems clear that even after performing a
measurement on one of the particles of an entangled pair, the companion
particle cannot be ascribed a reality in the sense of Einstein}. The
companion particle's quantum properties remain as unmeasured and as
`un-collapsed' as ever, though the result of a measurement if performed, in
the same direction, can be predicted with absolute certainty. Wave-function
collapse in the sense of Copenhagen interpretation and realization of an
outcome happens only during actually performed measurements and not as a
consequence of a measurement on a subsystem of an entangled system. (I will
argue in another paper that the results of the Popper's experiment \cite
{popper,kim,unni} support this view).

The solution presented here resolves the problem, pointed out by EPR, of
simultaneous reality of noncommuting observables. In fact the solution
denies any reality to an actually unmeasured system. This suggests that
there are physical systems in nature that are beyond the scope of the
intuitive definition of EPR reality, just as the Copenhagen school
maintained. The approach we have taken here gives predictions for
correlations which are exactly the same as that would be obtained from the
quantum wave-function and operators, without the apparent nonlocal influence
of one measurement on the other. The nonlocality apparent in entanglement
correlation in quantum mechanics is not an inherent feature, but a
conclusion forced on us when using a restrictive definition of physical
reality.

The same analysis works for particles entangled in other sets of variables
like momentum and coordinate, and energy and time. The results follow from
the fact that all these cases of two particle entanglement can be mapped on
to the spin-$\frac{1}{2}$ singlet problem with two-valued outcomes. An
experiment in which the particles entangled in momentum and position are
used, with double slits for each of the particles, the amplitudes are

\begin{eqnarray}
C_{1} &=&\frac{1}{\sqrt{2}}\exp (i\alpha k(x_{1}-x_{o})/2),  \nonumber \\
C_{1} &=&\frac{1}{\sqrt{2}}\exp (i\alpha k(x_{2}-x_{o})/2)
\end{eqnarray}
where $x_{1}$ and $x_{2}$ are the coordinates of the two detectors separated
by a space-like interval. $k$ is the wave vector and $\alpha $ is a scaling
factor for the angle subtended by the two slits at the detectors, source
etc. The factor $2$ dividing the angular variable comes from the mapping
with the spin-$\frac{1}{2}$ problem. The single particle data on either side
separately do not show any interference. The correlation function is 
\begin{equation}
U(x_{1},x_{2})=\cos (\alpha k(x_{1}-x_{2})/2)
\end{equation}
Probability for coincidence detection is 
\begin{equation}
P(x_{1},x_{2})=\cos ^{2}(\alpha k(x_{1},x_{2})/2)=\frac{1}{2}(1+\cos k\alpha
(x_{1}-x_{2}))
\end{equation}

This is the two photon correlation pattern with 100\% visibility, derived
assuming locality of probability amplitudes. This agrees with the quantum
mechanical prediction from the relevant two-particle wave function.

We have also constructed local amplitudes for the Hardy experiment \cite
{hardy} in which quantum mechanics predicts three particular zero joint
probabilities are one nonzero joint probability (the other possible joint
probabilities in the problem can be nonzero and are not relevant for the
demonstration of nonlocality). Local complex amplitudes that reproduce the
four relevant joint probabilities can be constructed easily. It is
impossible to achieve this if locality at the level of probabilities are
assumed, as in a local realistic theory.

Quantum entanglement swapping \cite{swap} is understood within this frame
work by noting that Bell state measurements choose subensembles of particle
pairs that show a particular joint outcome. Particles entangled
independently with the pair of particles that are subjected to the Bell
state measurement will show a joint outcome consistent with swapped
entanglement due to the correlation encoded in the internal variable. {\it %
But the Bell state measurement does not collapse the distant particle into a
definite state}. Yet all correlations are correctly reproduced. This has
important implication to the interpretation of quantum teleportation.

In summary, the long standing puzzle of nonlocality in the EPR correlations
is resolved. There is no nonlocal influence between correlated particles
separated into space-like regions. The solution has new physical and
philosophical implications regarding the nature of reality, measurement and
state reduction in quantum systems. Our approach shows that the EPR paradox
of simultaneous reality for noncommuting physical variables arise from their
restrictive definition of physical reality.

By restoring locality into the quantum measurements of entangled system and
removing the undesirable `spooky action-at-a-distance', one of Einstein's
deepest wishes is realized. But his desire for a tangible concept of reality
of unmeasured quantum systems does not look tenable.

\medskip

\noindent \noindent E-mail address: unni@tifr.res.in, unni@iiap.ernet.in

\end{document}